\documentclass[10pt,a4paper]{article}
\usepackage{mathrsfs}
\usepackage{fancyhdr}
\usepackage[top=30truemm,bottom=30truemm,left=20truemm,right=50truemm]{geometry}
\usepackage{indentfirst}
\usepackage{authblk}
\usepackage{graphicx}% Include figure files
\usepackage{bm,color}% bold math   \textcolor{red} ON
\renewenvironment{abstract}{\begin{quotation}{\normalsize }}{\end{quotation}}
\makeatletter
\newcommand{\email}[1]{\newcommand{\@email}{E-mail: #1}}
\renewcommand{\maketitle}{
\newpage\null
    \vspace{2em}
 {\LARGE\bfseries\noindent\ignorespaces\@title\par}
 \vspace{1em}%
 {\large\noindent\ignorespaces\@author\par}
 \vspace{2mm}
 {\normalsize\noindent\ignorespaces\@email\par}
\vspace{1em}
}

\renewcommand\@author{\ifx\AB@affillist\AB@empty\AB@author\else
      \ifnum\value{affil}>\value{Maxaffil}\def\rlap##1{##1}%
     \vspace{1mm} \AB@authlist\\\AB@affillist
    \else  \AB@authors\fi\fi}
\makeatother
\usepackage{amsmath,amssymb}
\title{Large amplitude collective dynamic beyond the independent particle/quasiparticle picture}
\author[1]{Denis Lacroix}

\affil[1]{Institut de Physique Nucl\'eaire, IN2P3-CNRS, Universit\'e Paris-Sud, F-91406 Orsay Cedex, France}
\email{lacroix@ipno.in2p3.fr}
\begin{document}
\maketitle

%%%abstract
\begin{abstract}
{\bf Abstract: }
In the present note, 
a summary of selected aspects of time-dependent mean-field theory is first recalled. 
This approach is optimized to describe 
one-body degrees of freedom. A special focus is made on 
how this microscopic theory can be reduced to a macroscopic dynamic for a selected set 
of collective variables. Important physical phenomena like adiabaticity/diabaticity, one-body dissipation or 
memory effect are discussed. Special aspects related to the use of 
a time-dependent density functional  instead of a time-dependent Hartree-Fock theory based on a bare hamiltonian are underlined.
The absence of proper description of complex internal correlations however strongly impacts the 
predictive power of mean-field. 
A brief overview of theories going beyond the independent particles/quasi-particles theory is given. Then, a special attention is paid for finite fermionic systems 
at low internal excitation. In that case, quantum fluctuations in collective space that are poorly treated at the mean-field level, are
important.  Several approaches going beyond mean-field, that are anticipated to improve the description of quantum fluctuations, 
are discussed: the Balian-V\'en\'eroni variational principle,  the Time-Dependent Random Phase Approximation and the recently 
proposed Stochastic Mean-Field theory. Relations between these theories are underlined as well as their advantages and shortcomings.   
\end{abstract}
%%%%%%%

%%%keyword
{\bf keywords:} Density functional theory, Correlations, Quantum fluctuations.
%%%%%%

\section{Introduction}
The time-dependent mean-field approach provides a unique tool to 
describe nuclear dynamic from small to large amplitude collective motion.
Microscopic time-dependent Hartree-Fock (TDHF)  based on zero 
range interaction \cite{Bon76} were introduced only four years after  the seminal work of Vautherin and Brink \cite{Vau72}
showing that Skyrme like interaction can be used to construct an energy
functional of the density (Energy Density Functional [EDF]). The EDF leads to an accurate 
description of selected static properties in nuclei \cite{Rin80,Bla86}.  Static  and dynamical mean-field theories, by considering directly 
single-particle degrees of freedom (DOF) interacting through a common mean-field, was a major breakthrough 
in nuclear physics and opened the perspective of a unique framework able to describe static, dynamical and 
thermodynamical nuclear properties. In particular it avoids the preselection of few collective DOF, generally necessary 
in macroscopic theories. Already at the very early stage of the application of Time-Dependent EDF, it was realized that the 
independent particle picture used in the theory leads to severe limitations. In particular, mean-field theory that is anticipated to be optimal for one-body DOF provides a rather poor description of two-body and higher-order degrees of freedom. Many aspects of nuclei, like shape evolution, relative distance of two nuclei during a collision, ... are associated to one-body collective observables. 
Two-body or higher order DOF are connected to fluctuations in collective space. The poor treatment of quantum fluctuations 
leads to a quasi-classical description of collective motion within nuclear mean-field theory. This has important consequences on the 
predictive power of this approach. TD-EDF will be particularly powerful to describe specific physical phenomena where quantum fluctuations do not play a decisive role. It is for instance able to nicely predict the mean collective energy of giant resonances \cite{Lac04}. It also provides a very reasonable description of the fusion process above the Coulomb barrier.  In recent years, we have seen an increasing interest in applying TD-EDF due to  the possibility (i) to perform symmetry unrestricted 
evolution, i.e. full three-dimensional calculations (see other chapter of the present book) \cite{Kim97,Sim01,Nak05,Mar05,Uma05,
Was08,Sim10,Sim12a} (ii) due to the possibility to use complete Skyrme functionals compatible with state of the art nuclear structure calculations (iii) more recently, the possibility to incorporate pairing correlations \cite{Ave08,Eba10,Ste11,Sca12,Has12,Sca13,Bul13}.  
With these improvements, many applications ranging from collective vibrations, fusion, deep inelastic collisions like multi-nucleon transfer or quasi-fission or more recently fission have shown that microscopic mean-field theories can provide important physical information. 
Unfortunately, in all cases, some 
shortcomings uncovering the poor treatment of quantum aspects in collective space recurrently shows up:
\begin{itemize}
  \item {\bf Life-time of giant resonances:} the  mean energy of giant resonances which are obtained using TD-EDF subject to small  
  perturbation 
  are usually properly described. This is usually not the case of its life-time. Indeed, collective motion are expected to be damped out 
  due to the coupling of this excitations with more complex internal degrees of freedom.  This coupling directly reflects 
  correlations inside the nucleus and is not properly treated in TD-EDF. As a consequence, 
  the spreading width of the nuclear response is usually 
  severely underestimated (see for instance \cite{Lac04}). Giant resonances associated to shape oscillations are associated to one-body degrees of freedom and therefore TD-EDF should a priori be an optimal approach for this process. While, as we will see, this is indeed the case for short time, other DOFs might affect the one-body dynamic for long time.  The long-time description of nuclear systems requires to incorporate the effect of complex DOFs that has been disregarded at the mean-field level, at least in an approximate way.   
  \item {\bf Underestimation of experimental fluctuations:} when TD-EDF was applied to deep inelastic collision, it was immediately realized that it can rather well describe the mean properties of the reaction partners like the average mass of the quasi-projectile or its 
  average total kinetic energy but completely fails to describe the width of experimental distributions. 
  From the extensive comparison between experiments and theory, two important conclusions 
  can be drawn. Since the average properties of collective variables are well describe and since important dissipation occurs in deep 
  inelastic collisions, TD-EDF properly incorporate dissipation in one-body space. More precisely, it accounts for the so-called one-body 
  dissipation effects that is associated to both the nucleon exchange process and to the reaction of single-particle states to a dynamical 
  shape deformation of the nucleus. Experimental fluctuations are directly linked  to fluctuations in collective space. These fluctuations 
  that are associated to two-body operators, are beyond the scope of the theory.  It should be noted that quantum fluctuations are 
  not zero in a mean-field theory and therefore this approach might have given a reasonable account for the fluctuations. Unfortunately, 
  nuclei are finite many-body quantum systems were large quantum fluctuations takes place and were a simple mean-field theory 
  strongly underestimates them. 
  \item {\bf Tunneling in collective space:} a last example of the limitation of TD-EDF is the complete absence of quantum tunneling in 
  collective space. As a consequence, two nuclei cannot fuse if their center of mass energy is below the fusion barrier energy.  Similarly, 
  the spontaneous fission process that requires the fission barrier transmission cannot be described.  
  \end{itemize}

These few examples illustrate the necessity in some cases to improve the microscopic description of small and large amplitude dynamics beyond the independent particle picture. In the last decades, several methods have been proposed to correct mean-field 
evolution. Most often, due to the increase of complexity compared to the original TD-EDF, these theories have been only 
applied to rather schematic models or even not been applied at all. Nevertheless, TD-EDF has now reached a certain level 
of maturity and with the constantly increasing computational power, the standardization of time-dependent methods beyond mean-field appears as the next 
natural step in the field. The aim of this article is to provide some guidance in the beyond mean-field approaches dedicated to 
many-body fermionic motion. 
Several reviews have been recently made on this subject either focusing on the inclusion of correlations \cite{Lac04}, or on 
the addition of new DOFs through a variational ansatz \cite{Sim12a} or the use of stochastic methods \cite{Lac14}. The goal of the 
present article is not to redo an exhaustive review but rather to present selected theories that might be of practical interest in the coming 
years focusing especially on collective aspects. First, some general aspects of mean-field are recalled. Specific connections between single-particle and collective DOFs are introduced. Finally, several theories that are expected to improve mean-field and to treat
quantum fluctuations in collective space are underlined.

\section{Mean-field approach and collective evolution} 

The connection between mean-field and selection of one-body DOFs is best illustrated starting from 
a variational principle 
\begin{eqnarray}
S = \int_{t_0}^{t_1} dt \left\langle  \Phi(t) \right|  i\hbar \partial_t - \hat H  \left|  \Phi(t)  \right\rangle,
\label{eq:varia}
\end{eqnarray} 
where $\hat H$ is the hamiltonian while the variation of the action is restricted to Slater determinants. Due to the Thouless theorem \cite{Tho60}, 
local variation in the Slater determinant Hilbert space can be written as:
\begin{eqnarray}
| \Phi + \delta \Phi \rangle & = & e^{\sum_{ij} \delta Z_{ij} \hat a^\dagger_i \hat a_j }   | \Phi \rangle.
\end{eqnarray}   
Said differently, the one-body DOF associated to the set of operators $\{ \hat a^\dagger_i \hat a_j \}$ are the generators of the local transformation between Slater determinants. This specific property in combination with the use of the variational principle  
(\ref{eq:varia}) insures that one-body DOFs fulfill the Ehrenfest evolution \cite{Fel00, Lac14}:
\begin{eqnarray}
i\hbar \frac{d \langle \hat a^\dagger_i \hat a_j \rangle }{dt} & = &  \langle \Phi(t) | [\hat a^\dagger_i \hat a_j  , \hat H] | \Phi(t) \rangle .
\label{eq:ehrenfest} 
\end{eqnarray}  

At this stage, many important comments can be made:
\begin{itemize}
  \item Equation (\ref{eq:ehrenfest}) is valid all along the dynamical path. 
  This does not mean that the one-body evolution is exact because there is no reason that 
  the exact time-dependent state remains a Slater determinant (SD) at all time. Starting from a SD state, due to 
  correlations, the exact many-body state will be a complex superposition of many SD. However, if the state 
  is initially prepared as a SD, the one-body density match the exact evolution for short time.     
  It is in that sense that the dynamic of one-body DOFs is optimized.
  \item Eq. (\ref{eq:ehrenfest}) is a priori only valid for single-particle operators. In particular, the Ehrenfest equation 
  is not fulfilled for the two-body density matrix components $\langle  \hat a^\dagger_i \hat a^\dagger_j  \hat a_k \hat a_l \rangle$ along the mean-field path. 
  So even for short time, two-body and/or higher order effects might be poorly treated.
  \item Since the expectation value in Eq. (\ref{eq:ehrenfest}) is taken over a Slater determinant, due to the applicability of the Wick theorem, the right hand side of (\ref{eq:ehrenfest}) becomes also a functional solely of the one-body density components. We then end-up with a closed set of equation that can schematically be written as: 
  \begin{eqnarray}
i\hbar \frac{d \langle \hat a^\dagger_i \hat a_j \rangle }{dt} & = &{\cal F} 
\left(\{  \langle \hat a^\dagger_k \hat a_l \rangle  \}\right) .  \label{eq:mfone}
\end{eqnarray} 
Denoting by $\rho_{ji} =  \langle \hat a^\dagger_i \hat a_j \rangle$ the one-body density matrix component. The above equation 
identifies with the standard TDHF evolution:
\begin{eqnarray}
i\hbar \partial_t \rho & = & \left[ h(\rho) , \rho \right] \label{eq:mf}
\end{eqnarray}
where $h(\rho)$ is the self-consistent mean-field hamiltonian.
\item The mean-field equation (\ref{eq:mf}) applies to more general situations where the system is not necessarily described by a Slater 
determinant but by a many-body density written as:
\begin{eqnarray}
\hat D & = & \exp\left(-\sum_{ij} Z_{ij} a^\dagger_i a_j + Z_0\right), \label{eq:dstat}
\end{eqnarray} 
where $Z_0$ insure that the density matrix is properly normalized. A new variational principle should be introduced 
starting directly from a density matrix instead of a trial wave-function. Such a variational principle has been proposed by Balian and 
V\'en\'eroni in the 80's \cite{Bal81,Bal84,Bal85}. For a recent comprehensive discussion see \cite{Sim12a}. The connection between the dynamical path and the selection of few DOFs is directly visible in this variational principle. Given a set of observables, denoted generically by ${\mathbf A} = \{ \hat A_\alpha \}$  and denoting by ${\mathbf A} (t)$ and $D(t)$ the observable and density matrix written in the Heisenberg representation. The action to minimize becomes:
\begin{eqnarray}
{\cal S} = {\rm Tr} \left( {\mathbf A}(t_f) \hat D(t_f)\right) - \int_{t_i}^{t_f} dt {\rm Tr} \left\{ {\mathbf A}(t) \left( \frac{d \hat D(t)}{dt} + \frac{i}{\hbar}
\left[ \hat H(t), \hat D(t) \right]\right) \right\} . \label{eq:bv}
\end{eqnarray} 
The action is varied by imposing the two boundary conditions $\hat D(t_i) = \hat D_i$, $ {\mathbf A}(t_f)= {\mathbf A}_f$. We see in particular from this variational principle, that the evolution will depend on the specific set of observables that are selected.  In particular, when the action is varied on the specific class of densities given by Eq.  (\ref{eq:dstat}) and the selected sets of relevant observables are one-body observables, i.e. $ {\mathbf A} = \sum_{ij} Z'_{ij} a^\dagger_i a_j$, the action becomes a functional of ${\mathbf Z}(t)$
and ${\mathbf Z}'(t)$, i.e.  ${\cal S}=   {\cal S}({\mathbf Z}(t), {\mathbf Z}'(t))$. the variation with respect to ${\mathbf Z}'(t)$ gives:
\begin{eqnarray}
i\hbar  {\rm Tr}  \left( \hat a^\dagger_i \hat a_j  \frac{d\hat D(t)}{dt}\right)& = & {\rm Tr} \left(  [\hat a^\dagger_i \hat a_j  , \hat H] \hat D(t) \right) ,
\label{eq:ehrenfest2} 
\end{eqnarray} 
that match also the exact Ehrenfest dynamics expect that the density is now restricted to (\ref{eq:dstat}). 
Due to the specific algebra of densities given by Eq. (\ref{eq:dstat}), Wick theorem can also be applied and the mean-field Eq. (\ref{eq:mf}) is recovered. This more general derivation applies in particular if the initial state is a statistical ensemble of independent particles.  As we will see below, another interest of the BV variational principle is that it might also be used in cases where the observable of interest are not necessarily one-body observables.   

\item{\bf Pairing correlation:}  In the last ten years, efforts have been made to incorporate pairing correlations in mean-field applications. While technically more involved, the Time-Dependent Hartree-Fock Bogoliubov (TDHFB) theory is formally a straightforward generalization of TDHF \cite{Bla86} where the independent particles picture is replaced by independent quasi-particles. Breaking the $U(1)$ symmetry associated to particle number conservation leads to non-zero value of the anomalous density components, $\kappa_{ij} = \langle a_j a_i \rangle$. TDHFB 
approach is optimal for a generalization of one-body "pseudo-observables" that can now be written as (using the notation of \cite{Rin80}):
\begin{eqnarray}
\hat A & = & \sum_{ij} Z^{11}_{ij} a^\dagger_i \hat a_j + \sum_{ij} \left( Z^{20}_{ij} a_j a_i +  Z^{20*}_{ji} a^\dagger_i a^\dagger_j \right).
\end{eqnarray} 
Breaking the particle number symmetry, is an elegant way to include part of the correlations, i.e. long-range pairing correlation, in the description. Still it is often classified as a mean-field theory where the one-body density $\rho$ and mean-field $h[\rho]$ are 
respectively replaced by the generalized density 
${\cal R}$ and generalized mean-field ${\cal H}({\cal R})$ \cite{Sim10}. In particular, including pairing does not solve any of the
shortcoming listed in the introduction. The recent systematic applications of refs. \cite{Sca13a,Sca14} clearly 
demonstrate the absence of damping in giant resonance. In addition, although TDHFB considers an enlarged set of 
relevant observables, their dynamics is still very classical. In the following, mean-field (MF) term will be used for approaches like TDHF 
or TDHF, while beyond mean-field (BMF) is reserved to approaches beyond the independent particles or quasi-particles.

\item {\bf Nuclear mean-field and density functional theory:} Last, I would like to mention an important aspect of the nuclear mean-field that has sometime been underestimated. 
In the present section, mean-field is introduced starting from a well-defined hamiltonian $\hat H$. This situation 
will be referred here as the "Hamiltonian case". Most often, the hamiltonian case can only serve as a guidance in the nuclear 
mean-field context. Indeed, what we call a mean-field approach is actually a density functional theory (DFT) where the energy 
is written as a functional of the one-body density matrix (normal or generalized) and where the parameters of the functional are directly adjusted to reproduce at best the nuclear properties. The direct fit to experiment 
demonstrates that the functional contains already at the so-called 
mean-field level complex many-body correlations. The merging of terminology like HF (TDHF) and EDF (TD-EDF) stems from a useful 
artifact of the nuclear mean-field 
where, most often, the EDF is constructed using the concept of effective interaction. It should however be 
kept in mind that the effective hamiltonian technique is actually just a practical way to construct an energy that becomes a functional of the density.  The difference between EDF approach and the Hamiltonian case becomes evident noting that the effective interaction vertices might themselves depend on the density of the state to which the effective hamiltonian is applied or noting that the interaction 
in the mean-field channel differs from the one in the pairing channel. This subtlety has always to be kept in mind when considering mean-field or beyond mean-field approaches for nuclei. For instance, it is safer to directly formulate the variational principle starting directly 
from the energy $E(\rho)$ instead of the hamiltonian $\hat H$. If pairing is neglected, one possible action to minimize would be:
\begin{eqnarray}
S = \int_{t_i}^{t_f} dt \left[ i\hbar \sum_\alpha  \langle \varphi_\alpha (t) | \partial_t | \varphi_\alpha (t)\rangle   - E[\rho(t)]\right] ,
\end{eqnarray}
where the single-particle state $| \varphi_\alpha \rangle$ are the occupied state, i.e. $\rho(t) = \sum_\alpha | \varphi_\alpha \rangle \langle \varphi_\alpha(t) | $
Then, the mean-field is directly defined in terms of the functional derivative of the energy $h[\rho] = \partial E[\rho]/\partial \rho$. 
     
  \end{itemize}
  
 \subsection{Macroscopic reduction of the mean-field dynamics}
 \label{sec:macro}
 
\begin{figure} \begin{center}
\includegraphics[width=12.0cm]{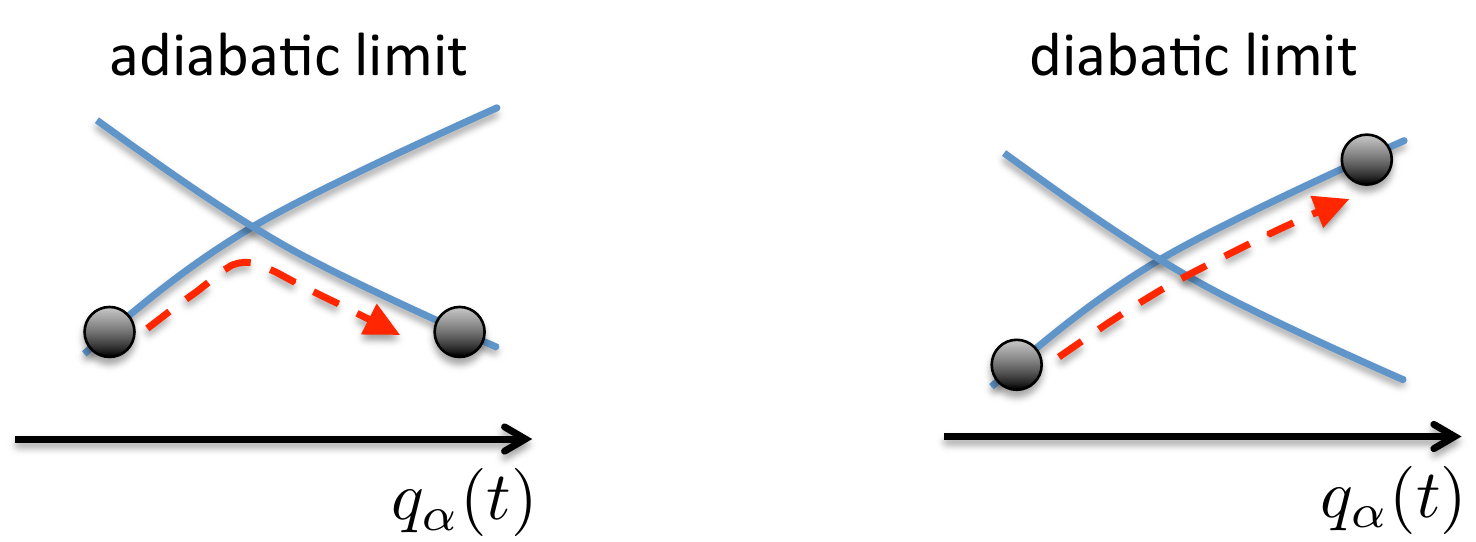}    
\caption{ \label{fig:adiab} Schematic illustration of the adiabatic (left) and diabatic (right) motions. In the adiabatic limit, in case of single-particle 
crossing, the lowest levels are always occupied. In the diabatic limit, the same levels remains occupied independently of level-crossing. This is for instance what happen if a simple scaling assumption is made for the wave-function. This scaling assumption is 
usually the starting point of hydrodynamical description.} 
\end{center} 
\end{figure}

The starting point of macroscopic approaches to small and large amplitude collective motion is to select few 
collective variables that are anticipated to play a dominant role in the physical phenomena under interest. For small amplitude 
vibrations, the operators generally correspond to multipole deformation operator \cite{Boh98}. When more than one nucleus is
concerned like in fusion and/or transfer reactions or larger amplitude deformation, a variety of collective variables are introduced: relative distance between nuclei, elongation, neck, mass/charge asymmetry... For simplicity, a single collective variable 
$q$ is considered here. In a simple markovian limit, the classical evolution of a macroscopic DOF might be schematically given by:
\begin{eqnarray}
\left\{
\begin{array}{ l}
\displaystyle \frac{d q}{dt}  =  \frac{p}{M}\\
\displaystyle \frac{d p}{d t}  =  -\left( \frac{\partial V(q)}{\partial q} \right)_{q} - \frac{1}{2} 
\left( \frac{\partial M} { \partial q} \right) p^2- \gamma(q). \dot q + \delta  p 
\end{array}
\right.
\label{eq:macro}
\end{eqnarray}       
In these equations, $p$ is the conjugate variable to $q$. $M(q)$ and $V(q)$ are respectively the collective inertia and potential energy landscape. The dissipation term $ \gamma(q). \dot q$ is generally introduced to account approximately for
the coupling between $q$ and other internal degrees of freedom, while $ \delta  p$ is a fluctuating term consistently introduced to 
fulfill the fluctuation-dissipation theorem requirement. In particular it might include possible thermal fluctuations. While simple approximations can be made, large effort is made to put as much as possible microscopic aspects in the different ingredients entering
into the macroscopic equations of motion \cite{Ada12,Ada14} . There are two main difficulties in the macroscopic framework (i) one should a 
priori pre-select  very few collective DOFs and therefore, one might miss important effects that are not anticipated. The validation of 
the hypothesis is generally made a posteriori by confronting results with experiments. (ii) The link between microscopic and 
macroscopic description 
can be made in general assuming specific approximation for the microscopic case, like very slow collective motion 
(adiabatic limit), or completely diabatic. An illustration of the two limits in the case of level crossing is given in Fig. \ref{fig:adiab}.   

One of the great advantage of dynamical mean-field approach is to avoid  arbitrariness in both (i) and (ii). 
Indeed, it does not pre-suppose that any one-body DOF plays a dominant role with respect to the others and do not assume 
a priori adiabaticity or diabaticity. Interesting connections between TD-EDF and the classical collective
dynamics given by Eq. (\ref{eq:macro}) can be made by considering specific limit of the evolution, either slow collective motion 
and/or  hydrodynamical evolution. The starting point of the adiabatic limit to TDHF is to assume that the one-body density takes the form \cite{Bri76,Gia76}
\begin{eqnarray}
\rho(t) &=& e^{i \chi(t)} \rho_0(t) e^{- i \chi(t)} ,
\end{eqnarray}
where both $\chi(t)$ and $\rho_0(t)$ are hermitian matrices. $\chi (t)$ plays the role of a velocity field that is common to all single-particle states. In the case of slow motion, $\chi(t)$ is small and the exponential can be expanded.  The variational principle 
(\ref{eq:varia}) then transforms as:
\begin{eqnarray}
S = \int_{t_0}^{t_1} dt  \left\{ - {\rm Tr}[\rho_0(t) \dot \chi(t)] - E[\rho_0(t), \chi(t)] \right\},
\label{eq:variaadiab}
\end{eqnarray} 
that is similar to the action in classical mechanics.  The connection with the macroscopic picture is then achieved by assuming
that the density $\rho(t)$ identifies with the density along a potential energy landscape obtained by minimizing the EDF constraining few collective DOF $q_\alpha$, i.e. $E(\rho_0(t)) = E(\{ q_\alpha(t) \})$. $\chi$ is then directly proportional to the velocities $\dot q_\alpha$. The Adiabatic limit of TD-EDF provides practical formula to evaluate the collective inertia and/or the collective potential. It however 
misses completely dissipative aspects since only few DOFs are explicitly followed in time. Possible coupling, even with other one-body DOFs, are therefore automatically neglected.  

\begin{figure} \begin{center}

 \includegraphics[width=8.0cm]{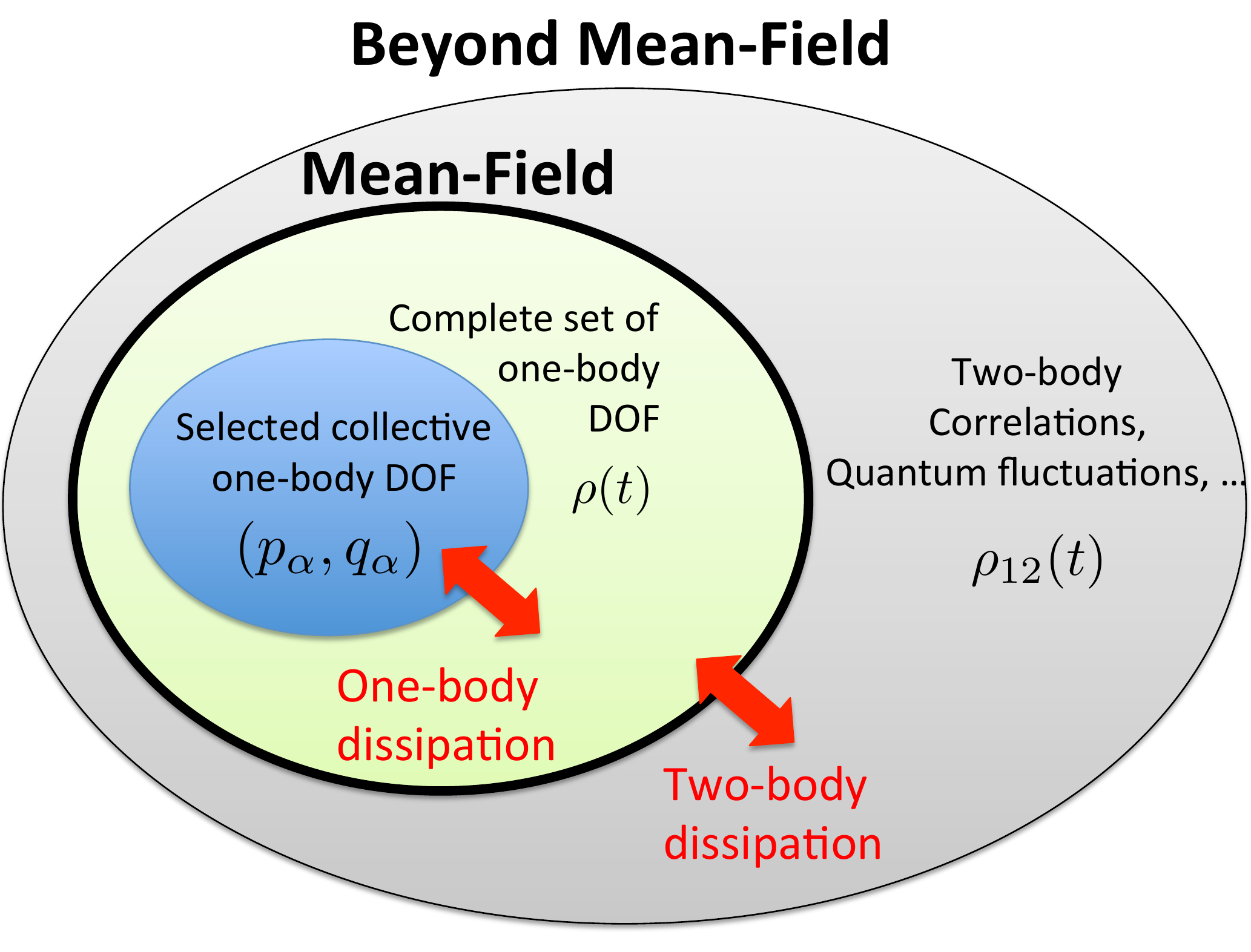}    
\caption{\label{fig:hier} A schematic representation of the different space of observables considered in the present article. 
The red arrows indicate the coupling between different spaces that will induce dissipation. Within the space of one-body DOF, focusing on few collective DOF, we expect a coupling between the collective variables and other one-body DOFs. This will lead to the one-body
dissipation mechanism that is already contained in microscopic mean-field theory. Going beyond mean-field can be considered as 
including two-body or higher-order correlations and/or quantum fluctuations in collective space. Treating the coupling between one-body
DOFs and new DOFs is at the heart of dissipation phenomena induced by internal correlations. } 
\end{center} 

\end{figure}

  The full TD-EDF includes this coupling and therefore should contains a priori the one-body dissipation mechanism (see the schematic picture given in Fig. \ref{fig:hier}). However, in the general 
case, the microscopic dynamic seems difficult to connect analytically to the transport coefficients of Eqs. (\ref{eq:macro}). Nevertheless, a method was developed in Refs.  \cite{Was08,Was09}. to extract directly the mass parameter, collective potential and dissipation coefficient for the relative distance during fusion.  The method can be described as follow. Given any one-body observable $\hat Q_\alpha$. Its dynamical evolution  can directly be obtained along the dynamical path from $q_\alpha (t) = 
{\rm Tr}(Q_\alpha \rho(t))$. For the relative motion, both the relative distance $R(t)$ and associated relative momentum $P(t)$
are known at all times. Assuming that their motion match the classical Eqs. (\ref{eq:macro}), one can invert these equations to get the mass, the potential and the dissipation kernel. Note that, since TD-EDF is a deterministic approach, there is no fluctuating contribution 
and $\delta p_\alpha=0$. 
The very reasonable collective potential extracted with this method and the compatibility of dissipation rate compared to empirical approaches again 
point out that the mean-field dynamic leads to a quasi-classical evolution in collective space.

 TDHF and/or TD-EDF is not expected to 
carry the information on two-body observables. Indeed, the theory is built up as a functional of the one-body density 
and its aim is solely to provide an accurate description of this quantity. 
\begin{figure} \begin{center}
\includegraphics[width=8.0cm]{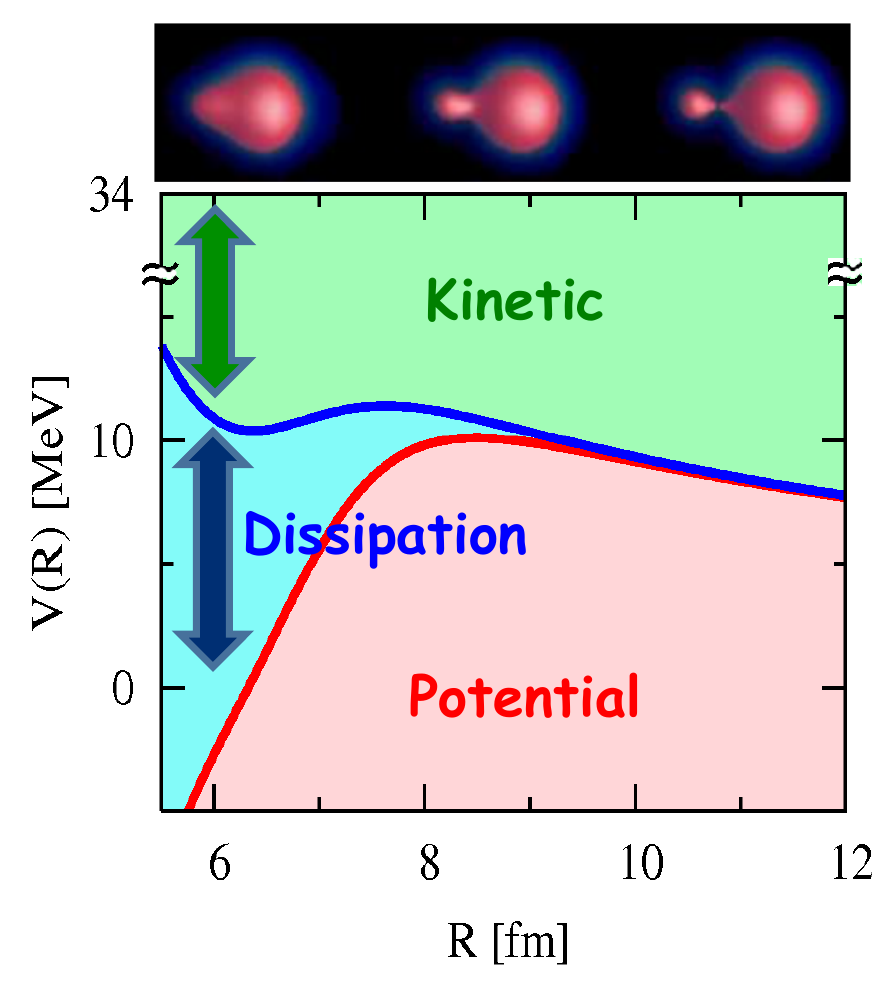}    
\caption{ \label{fig:vdiss} Illustration of the energy sharing during the fusion of $^{16}$O+ $^{16}$O assuming a simple macroscopic reduction of the mean-field dynamic. Adapted from Refs.  \cite{Was08,Was09}, the dissipated energy reported here is given by the simple formula 
$E_{\rm diss} = \int_0^t \gamma (q) \dot q^2(s) ds $.} 
\end{center} 
\end{figure}
Mean-field is therefore not built-up to carry the information on the
fluctuations in collective space since it belongs to two-body observables. Indeed, taking two one-body observables $\hat Q_1$ and $\hat Q_2$, their quantum correlations are given by:
\begin{eqnarray}
\sigma^2_{Q_1Q_2} (t)& = &  \frac{1}{2} \langle  \hat Q_1 \hat Q_2 + \hat Q_2 \hat Q_1 \rangle -   \langle \hat Q_1 \rangle 
\langle \hat Q_2  \rangle ,
\end{eqnarray}   
that obviously involves the two-body operators $\hat Q_1 \hat Q_2$ and $\hat Q_2 \hat Q_1 $. In a mean-field framework, these fluctuations become also a functional of the one-body density matrix and can be expressed as: 
\begin{eqnarray}
\sigma^2_{Q_1Q_2} (t)& = &  {\rm Tr}  \left[ Q_1 \rho(t) Q_2 (1-\rho(t))\right] ,
\end{eqnarray}
where the factor $(1-\rho)$ reflects the Fermi statistics of nucleons. It is worth mentioning that this fluctuations are a priori non-zero 
even at zero temperature. Therefore, the mean-field theory cannot completely be considered as a fully classical theory.  The situation 
we are facing with mean-field is similar to the case of coherent  states in quantum mechanics.  Coherent states are gaussian wave-packets whose properties is fully determined by the knowledge of their mean position $q$ and mean momentum $p$. Using coherent state in a variational principle leads to classical equation of motion for $(q(t),p(t))$. 
However, even if the quantum fluctuations are not completely neglected, 
most applications of mean-field theory to nuclear dynamics points out  that these fluctuations are severely underestimated compared to reality.  The treatment of correlations and/or fluctuations requires to go beyond the independent particle or quasi-particle picture \cite{Neg82}. 

\begin{figure}[htbp]
 \begin{center}
\includegraphics[width=10.0cm]{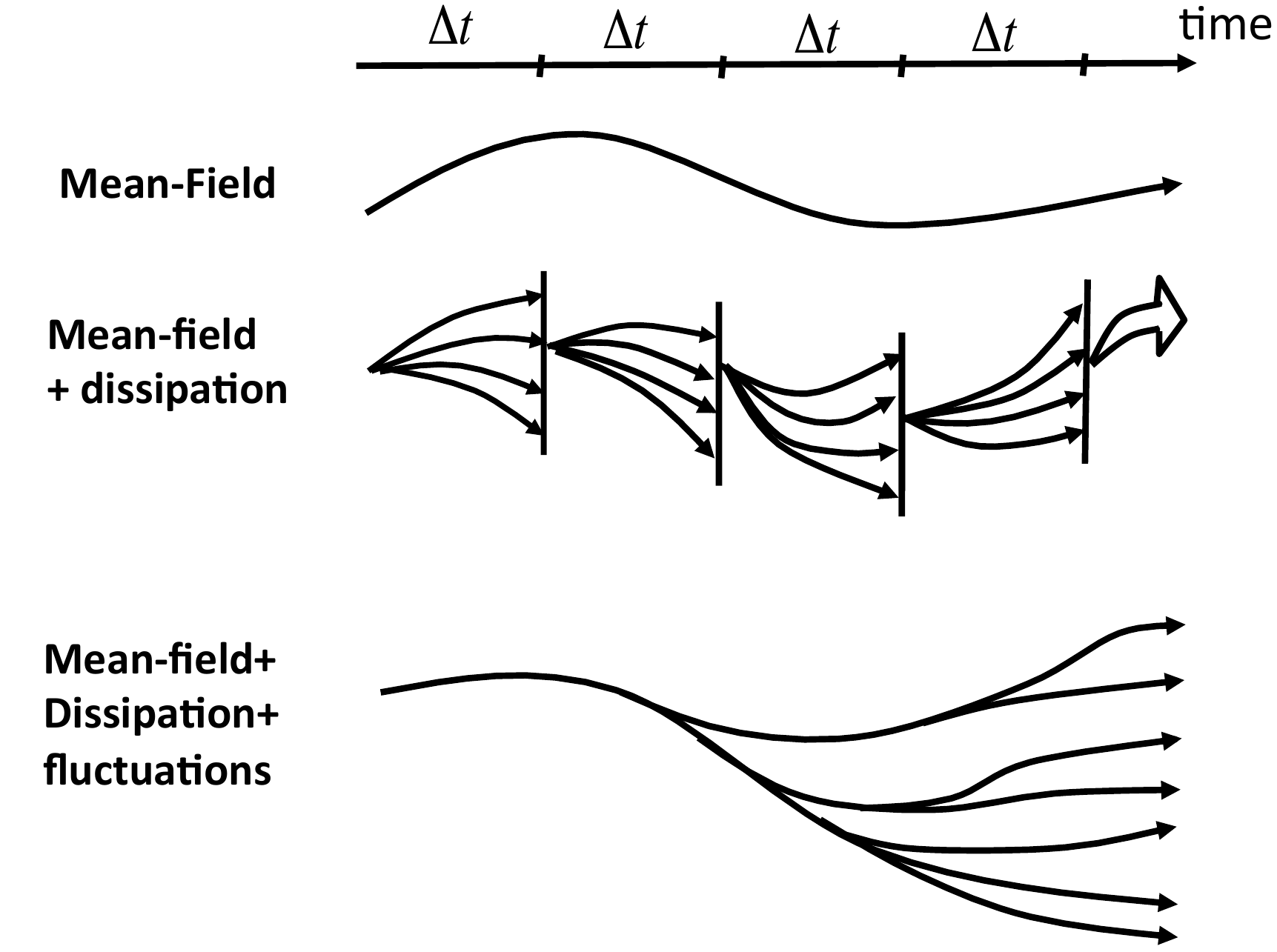}    
\caption{ \label{fig:scenario}Schematic illustration of the different scenarios  to include effect BMF keeping only the one-body density as relevant DOF.
Starting from the deterministic mean-field evolution, different degrees of sophistication can be used to include beyond mean-field effects. 
Keeping only the mean-field and the dissipative term ${\cal D} $ in Eq. (\ref{eq:gendisfluc}) lead to a new deterministic picture where 
the effect of complex DOFs is treated in an average way. Keeping both the dissipation and fluctuating term $\delta {\cal F}$ leads to 
a quantum Boltzmann approach where both dissipation and fluctuations are introduced consistently (bottom).
 A similar picture was used in ref. \cite{Ran92} to describe semi-classical methods dedicated to heavy-ion collisions around Fermi energy. }
\end{center} 

\end{figure}
\section{Beyond mean-field approach to large amplitude collective motion}

Beyond mean-field approaches is usually facing the difficulty to be much more involved in practice than the 
mean-field level. A common idea to BMF theory is to generalize mean-field by including more complex 
internal degrees of freedom linked to two-body or higher correlations. If we denote generically by ${\mathbf Q} = \{ \langle \hat Q_\alpha  \rangle\}$ the expectation values of the  
set of one-body DOFs and by ${\mathbf B} =  \{ \langle \hat B_i \rangle \}$ other degrees of freedom, the exact equation of motion can be formally written as:
\begin{eqnarray}
\left\{
 \begin{array}{l}
 \displaystyle
\frac{d}{dt} \langle \hat Q_\alpha \rangle   =  \langle [\hat Q_\alpha , \hat H ]\rangle = {\cal G}_\alpha ( {\mathbf Q} (t) ; {\mathbf B} (t)) \\
\\
\displaystyle\frac{d}{dt} \langle \hat B_i \rangle   =  \langle [\hat B_i , \hat H ]\rangle = {\cal G}_i ({\mathbf Q} (t) ; {\mathbf B}(t) ) 
\end{array}
\right. . \label{eq:coupled}
%( \langle {\mathfb Q } \rangle
%, \{ \langle \hat B_i \rangle \})  
\end{eqnarray}  
The main difficulty in solving this problem is the catastrophic explosion of number of DOF to follow in time. Indeed, supposing that the 
number of one-body DOF is $N$, the number of correlation matrix elements is $N^2/2$, the number of third moments is $N^3/2$, ... 
The set of equations given above contains the coupling between one-body DOF and more complex DOF that is responsible 
of two-body dissipation (see schematic picture \ref{fig:hier}).

Except very rare cases, specific strategies should be used to reduce the complexity of this problem. The great simplification of mean-field 
approximation is that all DOFs ${\mathbf B} $ becomes functional of the relevant DOFs ${\mathbf Q}$, i.e.  $\langle {\mathbf B} \rangle  = f(\langle {\mathbf Q} \rangle)$. Then, the second set of equations in (\ref{eq:coupled}) becomes redundant and we simply have $N$ coupled transport equations:
\begin{eqnarray}
\frac{d{\mathbf Q} }{dt}  = {\cal F} ({\mathbf Q}) .  \label{eq:mfq}
\end{eqnarray}  
This equation is nothing but Eq. (\ref{eq:mfone}) already shown in previous section. 

Three strategies commonly used to extend mean-field are summarized below:
\begin{itemize}
%\begin{description}
 \item[(i)] Following all DOFs in the coupled equations (\ref{eq:coupled}) is usually prohibitive. When possible, it is preferable to not follow them explicitly but correct the $\langle \hat Q_\alpha \rangle$  evolution by including the effect of other DOFs in an approximate way. The problem we face is similar to an open quantum system (OQS) problem where a system is coupled to a complex environment \cite{Bre02}. This similarity provides strong guidance to the N-body problem. The starting point to avoid the explicit treatment of the environment  DOFs is to first integrate the second equation in (\ref{eq:coupled}), leading to:
 \begin{eqnarray}
 \langle \hat B_i (t) \rangle   =   \langle \hat B_i (t_0)  \rangle + \int_{t_0}^t  {\cal G}_i ({\mathbf Q} (s) ; {\mathbf B}(s) ) ds . \label{eq:pert}
\end{eqnarray}
The appearance of the expectation value of $B$ at different time on the left and right hand side of this equation is similar to 
what happen in the Green function theory. If the coupling between different degrees of freedom is not too strong, 
a perturbative approach can be constructed from Eq. (\ref{eq:pert}). In many-body physics, an expansion a la 
Nakajima-Zwanzig (NZ) \cite{Nak58,Zwa60} is followed. Reporting the integral equation (\ref{eq:pert}) with some approximation 
in the first equation (\ref{eq:coupled}), lead to new equation of motion for the relevant DOF only, that could be written as:
\begin{eqnarray}
\frac{d{\mathbf Q}}{dt} = {\cal F} ({\mathbf Q}) + {\cal D} ({\mathbf Q})  + \delta {\cal F} ({\mathbf Q}). \label{eq:gendisfluc}
\end{eqnarray}   
 The first term in this equation is nothing but the mean-field. The second term, sometimes called dissipation kernel or 
 dissipator contains the accumulated effect of the coupling from the initial time $t_0$ to a given time $t$. It stems from the integral 
 in time in Eq.  (\ref{eq:pert}) and therefore depends on the full history of the system, i.e. on ${\mathbf Q}(s)$ with $t_0 \le s \le t$. This non-locality in time is sometimes called memory effect and/or non-Markovian effect. The importance of memory  is well known 
 especially in quantum systems \cite{Bre02} and has been for instance illustrated in Ref. \cite{Lac98}. It is worth mentioning that the 
 Nakajima-Zwanzig technique that leads to equations that are non-local in time for the relevant DOF has been recognized to not be the most efficient 
 way to treat open quantum systems. For instance, other techniques like time-convolutionless (TCL) seems more adequate \cite{Bre02}.
 In that cases, the equation of motion for relevant are time local and memory effects are incorporated in the transport coefficients. 
 This issue is largely unexplored in our field.
 
 The analogy between the N-body problem and OQS can be further exploited to obtain a Lindblad equation \cite{Bre02} from the dissipation kernel starting from Eq. (\ref{eq:gendisfluc}) \cite{Lac06}. Ultimately, the dissipation kernel can then be interpreted in terms of quantum jumps between independent particle states.  
 
 The third term $\delta {\cal F} ({\mathbf Q})$ contains the effect of initial correlations 
that are propagated in time through the mean-field. Invoking the absence of knowledge of this initial correlation this term is sometimes 
treated as a noise added to the mean-field. 
Comparing the original mean-field equation (\ref{eq:mfq}) and the new one, Eq. (\ref{eq:gendisfluc}) that includes eventually dissipation and fluctuation, the increase of difficulty  stems from the memory effect that is important in quantum mechanics and from the necessity to follow several trajectories instead of one if $\delta {\cal F} ({\mathbf Q})$ is treated as a fluctuating term. The different scenarios 
of beyond mean-field dynamics without treating complex degrees of freedom explicitly are shown in Fig. \ref{fig:scenario}.

  \item[(ii)] Another possibility is to add new degrees of freedom and try to solve a generalization of mean-field equation 
  in the form (\ref{eq:coupled}). Due to the number of DOFs (one-body, two-body, three-body, ...) it is just impossible to include all 
  of them and a specific truncation is required, for instance in the Time-Dependent Density Matrix approach (TDDM) \cite{Lac04}, 
  the two-body correlation matrix is followed in time while neglecting three-body correlations. The absence of a clear criteria to 
  truncate the hierarchy of equations jeopardize applications (see for instance Ref. \cite{Akb12}). Still few applications have shown that 
  correlations can be partially treated in this way (see \cite{Ass09,Toh15} and references therein).
  
  \item[(iii)] Finally, a third possibility is to consider more complex wave-functions allowing for a mixing 
  of single-particle states. For instance, if the aim is to re-quantize the motion in collective space. A possible method is to use trial state vector that is already a mixing of several independent particle or quasi-particle states.  Then,  the trial wave-function to be used in the variational principle can generically be written as:
   \begin{eqnarray}
| \Psi (t) \rangle &=& \int f({\mathbf Q}, t)  | \Phi({\mathbf Q}, t )\rangle  d{\mathbf Q} ,  \label{eq:gcm}
\end{eqnarray}    
where $\{  | \Phi({\mathbf Q}, t )\rangle \}$ stands for a set of eventually time-dependent quasi-particle vacuum. This general approach is 
called Time-Dependent Generator Coordinate Method (TD-GCM).
While this approach starts to be a standard tool for nuclear structure, very few discussions and applications have been made 
for the dynamical process (see for instance the discussion in ref. \cite{Goe80,Goe81} and application in  \cite{Gou05}). In particular, one of the extra difficulty is the doubling of collective space size compared to the static case, due to the necessity to include pairs of conjugated collective variables.  
%\end{description} 
\end{itemize}

The two strategies (i) and (ii) have attracted more attention in the last 30 years in the non-equilibrium context compared to the strategy (iii). However, strategy (iii) is the one that is now being recognize as one of the standard tool to extend mean-field and obtain spectroscopic information in nuclei. In the latter case, important discussions have been made recently to reconcile the GCM approach 
that is well defined starting from a true Hamiltonian \cite{Rin80} with the energy density functional theory where no Hamiltonian exists 
at the first place \cite{Dob07,Lac09,Ben09,Dug09}.

The TD-EDF theory is also strongly guided by the Hamiltonian case. Most often, if not always, final equations of motion in the three strategies are derived starting from a Hamiltonian and directly applied within the EDF context. The analogy 
with the Hamiltonian case should ultimately be taken with care and in particular, strategies (i), (iii) and (iii), 
when they are used in the nuclear context,  should be seen as methods to enrich the functionals. Indeed, in the TD-DFT context, the very notion of mean-field level and beyond mean-field level is a ill defined concept since already the approach should provide 
an exact description of the local density without invoking additional DOFs.  In the DFT context, the three strategies listed above 
should be regarded as follow:
\begin{itemize}
  \item The strategy (i) can be seen as a way to construct a richer functional of the one-body density matrix. In particular, 
  the appearance of time non-locality is not surprising in the DFT context where the functional is expected to depend on the history 
  of the local density \cite{Run84}. If the non-locality is important, the functional is called "non-adiabatic". This terminology should not be confused with the non-adiabaticity discussed in section \ref{sec:macro}.       
  \item The strategy (ii) should be interpreted as an alternative functional theory where the energy becomes a functional not only 
  of the one-body density but also of more complex degrees of freedom.   
  \item The strategy (iii) is certainly the one that is more difficult to understand in terms of a DFT. In Refs. \cite{Hup11,Hup12}, it has been shown that the many-body state $| \Psi  \rangle$ entering in Eq. (\ref{eq:gcm})  can be regarded as a specific trial state from which 
 the one-body and two-body densities are constructed leading to an energy functional of these quantities. We then end up with the schematic DFT scheme:
 \begin{eqnarray}
| \Psi \rangle & \longrightarrow & (\rho, \rho_{12}, \cdots ) \longrightarrow E(\rho, \rho_{12}, \cdots ).  
\end{eqnarray}   
The work in \cite{Hup11,Hup12} was dedicated to static properties and the application to nuclear dynamics within the EDF 
approach still remains to be formally clarified.
\end{itemize}  

In this section, a schematic presentation of beyond mean-field approaches is given trying to underlined some specific discussions 
related to the use of the Energy Density Functional approach. As we have seen above, the method used to extend mean-field is not 
unique and depends essentially on the physical process we want to include. Below, I concentrate on the specific 
description of finite many-body systems at low internal excitation. In that case, it is expected that the main difficulty 
is to treat quantum fluctuations beyond the independent particle approximation. Discussions on methods specifically dedicated 
to system at higher internal excitation can be found in \cite{Lac04,Lac14, Abe96}.

\section{Propagation of Quantal fluctuations in Beyond Mean-Field theories}

\begin{figure}[htbp]
 \begin{center}
\includegraphics[width=8.0cm]{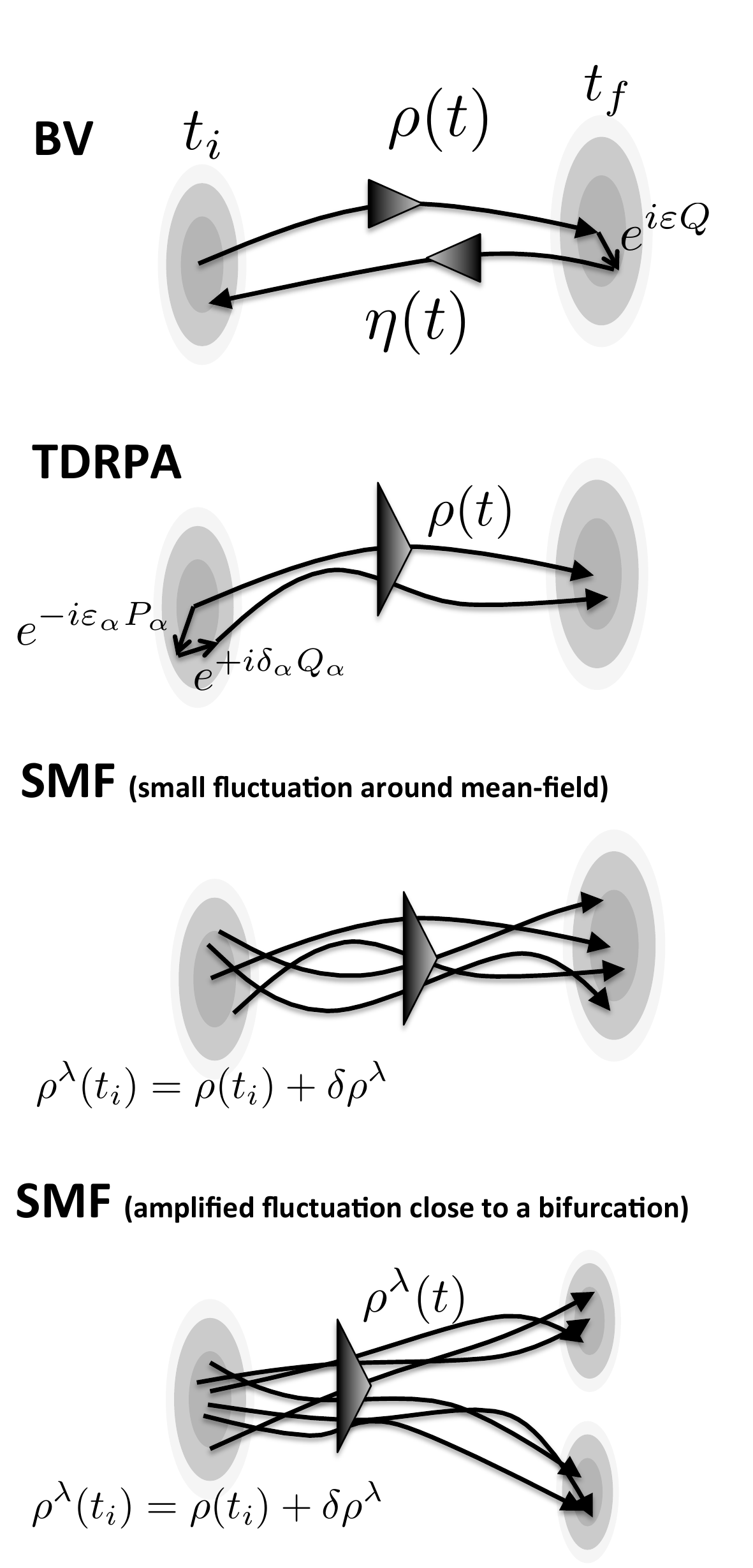}    
\caption{\label{fig:quantumfluc} Schematic illustration of the different approaches discussed in the text to describe quantum fluctuations beyond the mean-field theory (see text for more details).}
\end{center} 

\end{figure}

One of the deficiencies of mean-field transport theory is its inability to describe quantal fluctuations around the average 
properties in collective space.  This shortcoming is directly related to the quasi-classical nature of the collective evolution. 
At low internal excitation, this aspect is anticipated to be the most important effect beyond mean-field. The treatment of quantum fluctuations requires a priori to re-quantize the collective dynamics by considering a quantum wave-function in collective space, like 
in the TD-GCM (item (iii) above). This approach, involving a coherent mixing of many independent quasi-particles many-body states remains quite involved and has rarely been exploited in the dynamical context.  Alternative methods have been proposed that are trying to account for quantum fluctuations while keeping the simplicity of independent particle/quasi-particle states. Some of the approaches, depicted schematically in Fig. \ref{fig:quantumfluc} are briefly described in this section. 

\subsection{The Balian and V\'en\'eroni variational principle:} 
\label{sec:bv}

The variational principle proposed in the 80$^{th}$ by Balian and V\'en\'eroni, Eq. 
  (\ref{eq:bv}) can be used to derive time-dependent mean-field if the relevant degrees of freedom are restricted to one-body DOFs.   
  Generalization of TDHF can be obtained either by considering additional DOF, like two-body or higher correlations or by considering 
  more general many-body density matrix or both. There are essentially two difficulties in using the BV variational principle (i) if  
  complex density matrix are used, one looses the advantages of the the algebra of densities that are given by Eq. (\ref{eq:dstat}), 
  see for instance \cite{Flo89} (ii) the equations of motion obtained by adding more complex internal 
  DOFs become rapidly rather involved \cite{Bal81,Bal84,Bal85} and, as far as I know, have not been so far really integrated in time. 
  In order to include the fluctuations associated to a given one-body observable without facing the problem (i), BV proposed to optimize the set of observables given by $\hat A = e^{i\epsilon \hat Q}$ while $\hat D$ is still given by Eq. (\ref{eq:dstat}). Then, the expectation of $\hat A$ becomes the generating function of all moments of $\hat Q$. Still, the equations to be solved remain complicated. However, it was shown in Ref. \cite{Bal84,Bal92} that, in the specific case where the density identifies with a Slater determinant density,   the fluctuations associated to a given variable can be obtained by performing uniquely forward and backward mean-field evolution between initial and final times $t_i$ to $t_f$. The forward evolution is obtained from an initial one-body density $\rho(t_i)$ and leads to the density $\rho(t_f)$.
  Then a second evolution starts from a density $\eta(t_f) = e^{i\varepsilon Q} \rho(t_f) e^{-i\varepsilon Q}$ that is evolved backward to give $\eta(t_i)$ (see illustration in Fig.  \ref{fig:quantumfluc}). Finally, the fluctuations at final time $t_f$ that accounts partially for two-body effects are given by:
  \begin{eqnarray}
\sigma^2_Q (t_f) & = & \lim_{\varepsilon \rightarrow 0} {\rm Tr} \left(  \frac{\rho(t_i) - \eta(t_i)}{2 \varepsilon} \right)^2.
\end{eqnarray}
This approach can eventually be generalized to include more than one collective variable and also applies if the state has non-zero pairing. With the progress made to perform TD-EDF,  a renewal of interest is observed to extend the mean-field through the BV variational principle \cite{Bro08,Bro09,Sim11} (see also the contribution of C. Simenel in this book). In particular, it is expected that, 
in a regime where the backward propagation remains close to the forward evolution, the quantum fluctuations are corrected due to 
the Random-Phase-Approximation (RPA) correlation around the mean-field trajectory. However, there are situations where a small 
difference in the initial conditions is anticipated to lead to large deviations even over short time.  This is the case, when the system is close to a bifurcation point associated to a spontaneous symmetry breaking.  While this issue remains to be clarified, the application shown in ref. \cite{Bon85} seems to indicate that the BV variational principle is inadequate in that case. 

\subsection{ The time-dependent RPA theory}
\label{sec:tdrpa}

A closely related approach that has been proposed approximately at the same time as the BV method 
consists in building up directly a RPA like theory all along the mean-field trajectory \cite{Rei80,Rei84}. 
This is the so-called time-dependent RPA theory. In the RPA (or QRPA) approach, 
a correlated ground state is constructed starting from a static independent/quasi-particle vacuum.  This correlated state 
becomes then a vacuum for quasi-boson operators written as a coherent  superposition of particle-hole excitations \cite{Rin80}. 
In RPA, the collective energy landscape is assumed to be a harmonic and quantal fluctuations beyond mean-field can be extracted. In Ref. \cite{Rei80}, this approach has been generalized to include quantal fluctuations along the mean-field trajectory. Assuming that (i) the average trajectory 
is weakly affected by correlations and identifies with the TDHF trajectory (ii) that the quantal fluctuations remains small, a RPA theory
is constructed at each time. In this theory, a new state $| \Psi(t) \rangle $ 
is constructed that is a vacuum of  a set of quasi-boson operators $\{ \hat B_\alpha(t) \}$, i.e. $ \hat B_\alpha(t) | \Psi(t) \rangle =0$.
To make contact with the collective space evolution, it is convenient to write the quasi-boson operators as:
\begin{eqnarray}
\hat B_\alpha(t) & = & \hat P_\alpha(t) - i \lambda_\alpha \hat Q_\alpha(t) 
\end{eqnarray} 
where $\lambda_\alpha = 2 \langle \hat P^2_\alpha \rangle$. The operators $(\hat P_\alpha, \hat Q_\alpha)$ are conjugated operators
along the path. In particular, they defined specific trajectories around the average mean-field evolution through the transformation:
\begin{eqnarray}
| [\Phi + \delta \Phi] (t) \rangle & = & e^{i \sum_\alpha \varepsilon_\alpha \hat Q_\alpha -i \delta_\alpha \hat P_\alpha  }  | \Phi (t) \rangle
\end{eqnarray}  
where $(\varepsilon_\alpha, \delta_\alpha)$ are small parameters.

The equation of motion are obtained using the variational principle for the correlated state $| \Psi (t) \rangle$ leading to coupled equations between the state $| \Phi(t) \rangle$ that evolves through the standard mean-field assuming $(\varepsilon_\alpha = 0, \delta_\alpha=0)$ and the set of operators evolution $(\hat P_\alpha (t), \hat Q_\alpha (t))$, written in the Heisenberg representation. 
Technical details are given in ref. \cite{Rei80}. 
As in the RPA case, it could be shown that the state $ | \Psi(t) \rangle$ is a coherent mixing of two-particles two-holes excitation with respect to the independent particle state $| \Phi(t) \rangle$ and therefore contains additional correlations leading to enhanced collective 
fluctuations. Again, as in RPA, the correlated state do not need to be explicitly constructed thanks to the quasi-boson approximation 
\cite{Rin80}.  

It is finally interesting to note that some connection can be made with other approaches. Indeed, let us assume a practical situation where a set of DOF $(\hat Q_\alpha(t_i), \hat P_\alpha(t_f))$ are known at initial time, as well as the TDHF starting point $| \Phi (t_f) \rangle$. At all time, the operators $(\hat Q_\alpha(t), \hat P_\alpha(t))$ are defined through the implicit formulas:
\begin{eqnarray}
\hat P_\alpha(t) | \Phi(t) \rangle & = & ~i \lim_{\delta_\alpha \rightarrow 0 } \frac{ | \Phi_{\delta_\alpha} (t) \rangle - | \Phi (t) \rangle }{\delta_\alpha} \nonumber \\
\hat Q_\alpha(t) | \Phi(t) \rangle & = & - i \lim_{\varepsilon_\alpha \rightarrow 0 } \frac{ | \Phi_{\varepsilon_\alpha} (t) \rangle - | \Phi (t) \rangle }{\varepsilon_\alpha}
\end{eqnarray} 
where $| \Phi_{\delta_\alpha} (t) \rangle$ and $ | \Phi_{\varepsilon_\alpha} (t) \rangle$ are independent particle/quasiparticles state that evolves through their own mean-field with an initial condition that is slightly shifted either in $P_\alpha$ or $Q_\alpha$. This method is
illustrated in Fig.  \ref{fig:quantumfluc}. Again this method only applies if the deviation between the different states considered 
initially remains small during the time evolution.

\subsection{The stochastic mean-field theory}

One of the advantage of the two approaches depicted above 
is that they only requires to follow standard mean-field. The practical way 
that consists in following two mean-field trajectories that are close in collective space is anticipated to fail when large deviation 
will occur even if the initial conditions are very close from each other. This is the case for instance in quantum chaotic systems or close 
to a bifurcation point associated to the appearance of saddle points in collective space. Another important aspect, is the inability of mean-field to spontaneously break symmetries. Such symmetry breaking is of particular importance especially in quantum many-body systems and induces usually large quantum fluctuations in collective space.  

In recent years, a new approach \cite{Ayi08,Lac14}, called Stochastic Mean-Field (SMF) has been proposed to overcome these difficulties where also only mean-field evolution is needed.  The two main hypothesis of SMF are:
\begin{enumerate}
  \item The mean-field equation of motion of the density can be regarded as the classical or quasi-classical limit 
  of a many-body fermionic problem. As we have seen above, mean-field equation leads indeed to quasi-classical
  trajectories where many quantum effects are missing.  
  \item Quantum dynamic  and in particular quantum fluctuations  evolution can be simulated using classical trajectories 
  with a proper sampling of the initial collective phase-space. Note that such a classical mapping is a known technique to simulate quantum objects and might even be exact in some cases \cite{Her84,Kay94}. While a different method is used to treat the phase-space, 
  a similar strategy has been used in bosonic systems, the so-called truncated Wigner method \cite{Sin02}.
\end{enumerate}

In practice, SMF has been directly formulated in density matrix space. 
Let us assume that the aim is to improve 
the description of a system that, at the mean-field level and time $t_0$, is described by a density of the form:
\begin{eqnarray}
\rho (t_0)= \sum_i 
|\varphi_i (t_0) \rangle n_i \langle \varphi_i (t_0)|. \label{eq:densmf}
\end{eqnarray}
Then the mean-field evolution reduces to the evolution of the set of single-particle states 
\begin{eqnarray}
i\hbar \frac{\partial}{\partial t} |\varphi_i (t) \rangle = h[\rho] |\varphi_i (t) \rangle,
\label{eq:phitdhf}
\end{eqnarray}
while keeping the occupation numbers constant.

In the SMF approach, a set of initial one-body densities  
\begin{eqnarray}
\rho^{\lambda} (t_0)& = & \sum_{ij}  |\varphi_i \rangle \rho^{\lambda}_{i, j} (t_0) \langle \varphi_j |  \label{eq:ini}
\end{eqnarray}
is considered, where $\lambda$ denotes a given initial density. The density matrix components
$\rho^{\lambda}_{i, j}$ are chosen in such a way that initially, the density obtained by averaging over 
different initial conditions identifies the density (\ref{eq:densmf}).

It was shown in ref. \cite{Ayi08} that a convenient choice for the statistical properties of the initial sampling is 
\begin{eqnarray}
\rho^{\lambda}_{i, j} (t_0) = \delta_{ij} n_i + \delta \rho^{\lambda}_{i j} (t_0), \label{eq:mean}
\end{eqnarray}
where $\delta \rho^{\lambda}_{i j} (t_0)$ are mean-zero random Gaussian numbers while
\begin{eqnarray}
\overline{ \delta \rho^\lambda_{i j } (t_0)\delta \rho^{\lambda *}_{ kl }(t_0)} &=&  \frac{1}{2} 
\delta_{i  l }\delta_{j  k }\left( n^\alpha_i (1-n^\beta_j) + n^\beta_j (1 - n^\alpha_i)\right). \label{eq:fluc}
\end{eqnarray}
The average is taken here on initial conditions. In this approach, each initial condition given by 
Eq. (\ref{eq:ini}) is then separately evolved with its own mean-field:
\begin{eqnarray}
i\hbar \frac{\partial \rho^\lambda (t) }{\partial t} & = & [h[\rho^\lambda(t)], \rho^\lambda (t)],
\end{eqnarray} 
independently from the other trajectories.
%\begin{eqnarray}
%i\hbar \frac{\partial}{\partial t} |\varphi^{\lambda}_i (t) \rangle = h[\rho^{\lambda}] |\varphi^{\lambda}_i (t) \rangle,
%\label{eq:puresmf}
%\end{eqnarray}
%while keeping  the density matrix components constant.  
Therefore, the evolution 
along each trajectory is similar to standard mean-field propagation and can be implement with existing codes.
Two schematic illustrations of the stochastic mean-field method are given in bottom parts of figure \ref{fig:quantumfluc}.
Note that the SMF approach has been generalize to include pairing in ref. \cite{Lac13}.
The technical details and success of SMF approach have been recently reviewed in ref. \cite{Lac14} and only
specific aspects related to collective motion are given here. 

\subsubsection{Collective dynamic in SMF}

In the original mean-field approach, even if the theory is underestimating the quantum fluctuations compared to reality, 
the mean values and fluctuations of a given one-body collective observable $\hat O$ are still given by a quantum average, i.e. 
\begin{eqnarray}
\langle \hat Q \rangle & = & {\rm Tr} (Q \rho) , ~~~~~~\sigma^2_Q = \langle \hat Q^2 \rangle - \langle \hat Q \rangle^2= {\rm Tr} (Q \rho Q (1- \rho)). 
\end{eqnarray}
In SMF, along a given trajectory associated to $\rho^\lambda(t)$, the observable evolution is given by:
\begin{eqnarray}
Q^\lambda (t) & = & {\rm Tr} (Q \rho^\lambda(t)).  
\end{eqnarray} 
The very notion of quantum average is lost and the variable $Q^\lambda (t) $ is now considered as a classical object. 
For instance, fluctuations are estimated using the classical average over different trajectories, i.e. :
\begin{eqnarray}
\Sigma^2_Q & = & \frac{1}{N_{\rm traj}} \sum_\lambda  Q^\lambda (t) Q^\lambda (t)  - \left( \frac{1}{N_{\rm traj}} \sum_\lambda  O^\lambda (t)  \right)^2 \nonumber \\
&\equiv& \overline{Q^2} - \overline{Q}^2  
\end{eqnarray}  
where $N_{\rm traj}$ is the number of trajectories.
It is noteworthy that the sampling of initial conditions made under the two constraints given by  Eqs. (\ref{eq:mean}) and (\ref{eq:fluc}) insures  
that the classical average  $\Sigma^2_Q$ 
made over the initial conditions on SMF match the quantum average $\sigma^2_Q$ of the original quantum mean-field 
at initial time.

\subsubsection{SMF with small deviation around the mean-field evolution}

One direct proof that a classical mapping of mean-field theory with appropriate sampling of initial conditions can grasp effects 
beyond the independent particle or quasi-particle picture, is the possibility to make connection with the BV and/or TDRPA when 
trajectories remain close to the average mean-field evolution all along the trajectory (third panel case in Fig. \ref{fig:quantumfluc}).  In Ref. \cite{Bal85}, it has been shown that, assuming small deviation with respect to the mean-field evolution, the BV variational principle corrects the quantum fluctuations by including the TD-RPA correction. Then the quantum fluctuations at final time are given by:
\begin{eqnarray}
\sigma^2_Q & = & {\rm Tr} \left[  Q(t_i) \rho(t_i) Q(t_i) (1-\rho(t_i)) \right] \label{eq:tdrpa1}
\end{eqnarray} 
where $Q(t_i)$ is the observable that is propagated backward from $t_f$ to $t_i$ through the time-dependent RPA equation 
with initial condition $Q(t_f) = Q$. In \cite{Ayi08}, assuming small fluctuations, it was also shown that the SMF theory leads also 
to Eq. (\ref{eq:tdrpa1}) showing that it can describe at least TDRPA correlations beyond mean-field. 
This was verified explicitly for the Lipkin model case \cite{Lac12} when the collective energy landscape is harmonic. In that case, both the BV/TDRPA theory \cite{Bon85} and SMF gives identical results that can hardly be distinguished from 
the exact solution.  

The implementation of SMF is however a priori more demanding than the practical solution of the BV approach described 
in section \ref{sec:bv} or \ref{sec:tdrpa} since it requires to consider enough trajectories to reproduce the initial phase-space.
However, similarly to the BV or TDRPA, assuming that all trajectories remains close to the average mean-field (TDHF) evolution, the 
explicit sampling of trajectories can be avoided.  Then, the numerical cost to obtain corrections to mean-field  within SMF becomes 
of the same order as for the original mean-field. The simplified implementation of SMF can be summarized as follow. Starting from
the mean-field equation, written in a generic way as in Eq. (\ref{eq:mfq}):
 \begin{eqnarray}
 \frac{d\rho^\lambda(t) }{dt}  = {\cal F} (\rho^\lambda (t)),
\end{eqnarray}
and focusing on a specific variable $Q$, following the same strategy as for TDHF (Eq. (\ref{eq:macro})) a macroscopic reduction 
of SMF can be made leading to an equation of motion:
\begin{eqnarray}
\frac{d Q^\lambda}{dt} & = & F(Q^\lambda (t)) + \nu (Q^\lambda(t), t) + \delta Q^\lambda(t)  
\end{eqnarray}
where $F$ is the driving force associated to the collective potential. $\nu$ is the dissipative kernel that contains 
on-body dissipation effect. Both $F$ and $\nu$ are already present in the mean-field level. $ \delta Q^\lambda(t) $ is the additional fluctuating term that stems from the initial sampling. It is a priori non-Markovian.
Assuming that all trajectories remain close to the average trajectory and 
that this trajectory is identical to the mean-field one, one can eventually approximate:
\begin{eqnarray}
F(Q^\lambda (t)) & \simeq &  F(\langle Q \rangle ),~~~
 \nu (Q^\lambda(t), t)  \simeq    \nu (\langle Q \rangle, t)
\end{eqnarray}
where $\langle Q \rangle$ is the expectation value associated to the standard time-dependent evolution. With this simplification, 
the average evolution of the fluctuation in SMF $\Sigma_{Q}(t)$ contains an additional diffusion term associated to the classical fluctuations $\overline { \delta Q^\lambda(t) \delta Q^\lambda(t)}$. For a given observable, the  estimate   of this quantity is far from being straightforward. However, using the Wigner transform and assuming no memory effect (markovian limit), an approximate method has been proposed in Ref. \cite{Ayi09} and further applied in \cite{Was09b,Yil11, Yil14} to heavy-ion collisions. 
The great interest of this approach is that corrections beyond mean-field are obtained directly from the mean-field evolution. 
Very recently, it was shown that 
a fully quantal treatment including memory effect can be made \cite{Ayi15} opening new perspectives.  

\subsubsection{SMF with large deviation around the average}

The above simplified method will not work if the densities along each trajectory starts 
to strongly deviate from the average. An illustration of such a case is given in the bottom part of 
figure \ref{fig:quantumfluc}. Then, a direct sampling of the initial phase-space followed by a set of time-dependent 
mean-field trajectory is unavoidable. Up to now, the SMF theory in its full glory has been applied to rather simple 
models.  The application to the Lipkin model \cite{Lac12} was the first attempts and has shown that not only 
the SMF theory is equivalent to the TDRPA for harmonic motion around the ground state but also that it works in situations 
where other approaches fails, for instance to describe the non-equilibrium dynamic close to a spontaneous 
symmetry breaking associated to a quantum phase-transition. 
In ref. \cite{Lac13}, this aspects has been used specifically to take advantage of the $U(1)$ symmetry breaking 
associated to particle 
number conservation, leading to a generalization of SMF including pairing. The new theory was validated using the pairing 
hamiltonian. The capability of SMF has then been further demonstrated with the Hubbard model, where fast non-equilibrium 
process can occur after a quantum quench. In that case, the approach was competitive with state of the art non-equilibrium 
Green Function theory \cite{Lac14a}. The successes of SMF are very promising. Still many aspects needs to be clarified:
 \begin{itemize}
 
  \item {\bf Predictive power:} In different applications of SMF, it was systematically observed that this theory can provide
   a very good approximation beyond mean-field in the weak coupling regime, i. e. when the residual interaction is weak. 
  In some cases \cite{Lac12,Lac13}, the SMF results cannot be distinguished from the exact solution.  When the strength of the two-body interaction increases, usually the SMF approach is rather good over a certain time and then, starts to deviate from the exact 
  evolution. In the Hubbard model case, in the strong interaction regime, it was observed that the SMF fails to reproduce 
  the dynamics even at rather short time scale. Up to now, the approach has been tested in a rather empirical way and its 
  success can only be tested comparing with the exact solution. In the near future, it would be desirable to seek for some criteria
  that would allow to anticipate when the theory would have a good predictive power.      
  
  \item {\bf Symmetry breaking:} A clear interesting aspect of the approach is to allow for initial densities $\rho^\lambda$ 
  that can break some symmetries that are respected by the initial average density $\rho(t)$.  In a physical system many symmetries 
  can be broken. In practice, the density $\rho^\lambda$ can be chosen to break all possible symmetries including $U(1)$ symmetry 
  or can break only selected symmetries while some others are respected. As a consequence, the SMF theory is not unique and can 
  be implement at various levels of symmetry breaking.  Obviously, the more symmetry are broken, the more demanding is the calculation. 
  The results of SMF strongly depend on the symmetries that are initially broken. For instance, in the Hubbard model, a better result
  is obtain if the spin up/spin down symmetry is broken compared to the case where it is not broken. On opposite, no improvement was observed if the $U(1)$ was broken compared to the case where it was not. 
   Again, the choice of breaking or not of some symmetries is rather empirical and requires a priori some physical intuition.
  \item {\bf Gaussian sampling of the phase-space:} In the original SMF theory formulation, only the first and second moments of the initial density fluctuations are constrained to match with the initial quantum fluctuations. This hypothesis, that implicitly 
  assume a Gaussian initial distribution of the density matrix has been recently questioned in an exploratory study \cite{Yil14b}.
  Using a more realistic sampling of the initial phase-space, it was shown that the SMF dynamics can be further improved, especially for larger coupling and longer time. This clearly points out that the approach and its predictive power is largely unexplored.     
  A key challenge to further progress is to have an efficient method able to provide a realistic phase-space in complex 
  many-body fermionic systems.  
  \item{\bf Correlated initial state:} Again in \cite{Yil14b}, it was also shown for the first time that the SMF can be applied even if the 
  initial state is not an independent particle system but contains correlations. The key of the success was the realistic sampling of the initial
  phase-space. 
\end{itemize}

\section{Summary}

After the first application in nuclear physics of time-dependent mean-field based on Skyrme effective 
interaction, it was realized very rapidly  that this approach should be extended to include beyond 
mean-field effects. During many years after this first application, a great number of famous scientists have proposed a variety
of approaches that includes different effects \cite{Neg82}. Most of the frameworks turn out to be so complex that they 
have never been applied so far. One of the difficulty is the quantum nature of nuclei. In view of this complexity, some 
theories where applied using a semi-classical approximation \cite{Abe96,Cho04}. In recent years, great progress have been made 
in the application of TD-EDF to the nuclear many body problem. This has renewed also the interest of developing quantum
transport theories beyond mean-field. In the present notes, an overview of several techniques that incorporate some 
effects beyond the independent particle/quasi-particle approaches is given.

In the second part of this article, a focus is made in finite many-body systems at low internal excitations.
In that case, quantum fluctuations in collective space are expected to play an important role. Mean-field 
approaches based on one-body density matrix are known to poorly describe quantum collective fluctuations.
Theories that can be regarded as practical tools to improve mean-field in the context of large amplitude 
collective motion are described: namely the Balian-V\'en\'eroni variational principle,  the Time-Dependent Random Phase Approximation and the Stochastic Mean-Field theory. Their advantages and drawbacks are discussed. Among them, the Stochastic Mean-Field 
approach seems to provide a versatile tool able to describe both the collective motion in the harmonic limit as well as
strong non-equilibrium process occurring close to a bifurcation point or a quantum phase-transition.  Last, I would like to mention 
that the present discussion is by no way exhaustive and the future of transport theories applied to finite 
quantum many-body systems is largely open to new ideas.

\section*{Acknowledgment}  
I would like to warmly thank all my collaborators for many fruitful discussion on 
collective dynamics: G. Adamian, N. Antonenko, S. Ayik, D. Gambacurta, G. Hupin, K. Washiyama, G. Scamps, 
C. Simenel, Y. Tanimura and  B. Yilmaz.

\end{document}